\documentclass[%
 reprint,
superscriptaddress,
 amsmath,amssymb,
 aps,
pra,
 floatfix,
]{revtex4-2}
\usepackage[utf8]{inputenc}
\usepackage[T2A]{fontenc}
\usepackage{amsmath,amssymb,amsfonts}
\usepackage{graphicx}
\usepackage{physics}
\usepackage{bm}
\usepackage{float}
\usepackage{algorithm}
\usepackage{algpseudocode}
\usepackage{braket}
\usepackage{color}
\newcommand{\bs}{\boldsymbol}

\begin{document}

\preprint{APS/123-QED}
\title{Simulation of boson sampling with optical feedback}
\author{Yu.A. Biriukov}
\altaffiliation[e-mail: ]{biriukov.ia18@physics.msu.ru}
\affiliation{
 Quantum Technology Centre and Faculty of Physics, M.V. Lomonosov Moscow State University, 1 Leninskie Gory, Moscow 119991, Russia
}

\author{I.V. Dyakonov}
\affiliation{
 Quantum Technology Centre and Faculty of Physics, M.V. Lomonosov Moscow State University, 1 Leninskie Gory, Moscow 119991, Russia
}
\affiliation{
Russian Quantum Center, 30 Bolshoy bul'var building 1, Moscow 121205, Russia
}

\date{\today}% It is always \today, today,
             %  but any date may be explicitly specified

\begin{abstract}
This work presents a theoretical model of boson sampling with optical feedback, in which a subset of the interferometer's output modes is looped back into the input modes. If the bosons are injected periodically into the input modes of the interferometer and optical feedback lines' length match the period of injection, it allows for interference between bosons injected at the consequent time iterations. We propose several methods methods for computing the output photon distributions in both output spacial and temporal modes, including not only standard spatiotemporal mode-unfolding technique, but also the Kraus-operator formalism, and a correlation-tensor-based approach. The two latter approaches help us to reveal that for random interferometers this system evolves to a unique stationary state over time. Because of the existence of the stationary state, we introduce new computational problem \textit{Stationary Distribution Boson Sampling} which appears to be harder than conventional boson sampling problem and contains it as a special case when there are no optical feedback lines.   
\end{abstract}

%\keywords{Suggested keywords}%Use showkeys class option if keyword
                              %display desired
\maketitle

\section{Introduction}
Boson sampling, introduced by Aaronson and Arkhipov~\cite{aaronson2011computational}, is one of the most promising candidates for demonstrating quantum computational advantage using linear-optical systems. The task consists of sampling from the probability distribution of $N$ indistinguishable photons at the output of an $M$-mode unitary interferometer, where individual outcome probabilities are proportional to the squared permanent of submatrices of the transfer matrix $U$~\cite{scheel2004permanents}. Because computing the permanent is \#P-hard~\cite{valiant1979complexity}, a scalable photonic boson sampler is widely believed to outperform any classical computer for sufficiently large instances~\cite{aaronson2011computational}.

The original proposal has inspired a rapid development of experimental platforms, starting with small-scale demonstrations using bulk~\cite{broome2013photonic,spring2013boson} and integrated~\cite{crespi2013integrated,spagnolo2014experimental} optics. Simultaneously, theoretical efforts have focused on classical simulation algorithms for validation purposes~\cite{neville2017classical,clifford2018classical}, the robustness of boson sampling to small implementation errors~\cite{arkhipov2015robust}, have proposed closely related variants such as scattershot and Gaussian-state boson sampling~\cite{lund2014gaussian, bentivegna2015experimental} and on extensions of the conventional boson sampling model that enrich its computational power for quantum simulation tasks~\cite{gonzalez2016quantum,arrazola2023quantum}. Notable examples include the simulation of molecular vibronic spectra and Franck--Condon factors~\cite{huh2015boson,sparrow2018simulating}, non-adiabatic chemical dynamics~\cite{fingerhuth2024quantum}, quantum walks in disordered media~\cite{caruso2016fast}, and many-body bosonic thermodynamics~\cite{arrazola2023quantum,zhang2024quantum}.

A particularly fruitful direction is the introduction of temporal degrees of freedom \cite{motes2015implementing, motes2014scalable, he2017time}. By allowing photons to circulate through the interferometer multiple times—either via optical delay lines or active feedback loops—one effectively creates a time-multiplexed architecture in which the evolving quantum state is preserved between temporal iterations. Such recurrent linear-optical networks have been shown to simulate discrete-time quantum walks~\cite{crespi2013integrated,valle2014two}, non-Markovian dynamics~\cite{caruso2016fast} Moreover, feedback-enabled boson samplers can dramatically reduce the physical resource overhead required to demonstrate quantum advantage, as the same spatial modes can be reused across many time steps~\cite{zhou2022timestamp}.

\begin{figure}[ht]
    \centering
    \includegraphics[width=\linewidth]{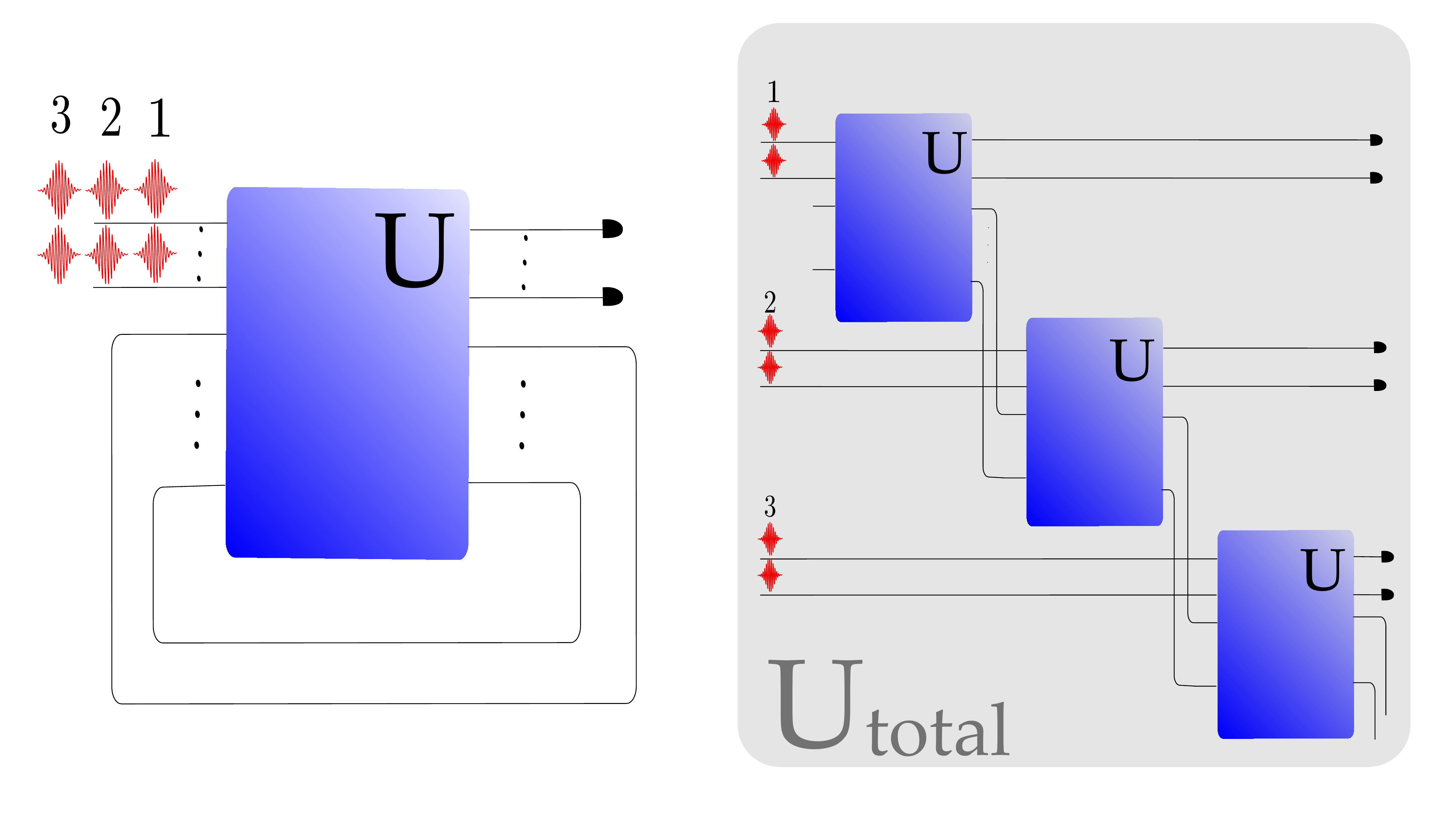}
    \caption{(Left) General scheme of a boson sampler with optical feedback channels and transfer matrix $U$. (Right) The same interferometer, but with all spatiotemporal modes mapped to consider the system as a standard boson sampler with total transfer matrix $U_{total}$.}
    \label{fig:mapping}
\end{figure}

In this work we propose and systematically study a boson-sampling architecture with optical feedback, in which a subset $L$ of the $M$ output modes of a unitary interferometer is routed back to the input ports, while fresh photons are continuously injected into the remaining $\left(M-L\right)$ external modes (Fig.~\ref{fig:mapping}). This architechture was proposed \cite{zhou2022timestamp}, more detailed analysis and experimental validation of the device was conducted in \cite{biriukov2025experimental} The looped modes thereby act as a short-term quantum memory that retains coherence and photon-number information across successive iterations. Unlike standard single-pass boson sampling, where the input state is reinitialised at every experimental shot, the present scheme supports genuine temporal evolution of a mixed many-photon state with variable particle number.

To expand the results mentioned in previous works that tried to treat a  boson sampler with optical feedback as a convensional boson sampler by introducing spacio-temporal modes, we develop several complementary theoretical and numerical frameworks for modelling this dynamics: (i) evolution of the partial density matrix in the looped modes, (ii) a Kraus-operator description that naturally incorporates photonic losses, and (iii) a correlation-tensor formalism that permits efficient reconstruction of both transient and stationary states. Using the Kraus and tensor approaches we prove that, for almost all random unitary matrices, the system possesses a unique stationary state, confirming that photons do not accumulate indefinitely in the feedback loop. Numerical simulations illustrate the stabilisation timescale, the structure of the stationary density operator (which is generally neither thermal nor coherent), and the accuracy of density-matrix reconstruction from limited sets of correlation functions.

The Kraus-operator formalism introduced here extends the scope of photonic boson sampling to the simulation of open quantum-system dynamics. At the same time, the precise computational complexity of classically simulating such feedback-driven architectures remains an open question: this is no longer the standard single-pass boson-sampling task for which the usual \#P-hardness arguments directly apply. Moreover, some very simple looped schemes admit an efficient tensor-network description~\cite{markov2008simulating,orus2014practical} and can therefore be simulated efficiently on a classical computer. In this work we aim to clarify where, in the space between these two extremes, the proposed looped architecture is located.

The developed methods are directly applicable to near-term integrated photonic experiments and pave the way for demonstrating quantum computational advantage in a time-multiplexed setting with substantially fewer spatial modes than required by conventional single-pass designs.

\section{Problem Statement}

Consider an $M$-mode linear optical circuit with a unitary transfer matrix $U$. Let $L$ output modes be connected back to the inputs, as shown in Fig.~\ref{fig:mapping} on the left. It is assumed that the loop length is such that photons leaving the circuit arrive at the next iteration simultaneously with photons at the external inputs. At each iteration $N$ photons are injected into the external inputs.

It is necessary to compute the photon distributions in the detectable modes or samples from this distribution at a given iteration of the time evolution. Furthermore, if the system reaches a stationary state, samples in the detectable modes must be computed for this case. Separately, the quantum state established in the looped modes needs to be studied.

\section{Structure of the State Space}

\subsection{Density Matrix}

The states of photons in the $M$ modes are described in the Fock space $\mathcal{H}_M$, which is the direct sum of subspaces with different photon numbers:
\[
\mathcal{H}_M = \bigoplus_{n=0}^{N_{\text{max}}} \mathcal{H}_M^{(n)},
\]
where $\mathcal{H}_M^{(n)}$ is the subspace of states with $n$ photons in $M$ modes, and $N_{\text{max}}$ is a sufficiently large number to encompass all possible multiphoton states. The dimensions of these subspaces are given by binomial coefficients:
\[
\dim \mathcal{H}_M^{(n)} = \binom{M + n - 1}{n},
\]
and the total dimension of the space $\mathcal{H}_M$ is:
\[
\dim \mathcal{H}_M = \sum_{n=0}^{N_{max}} \dim \mathcal{H}_M^{(n)} = \binom{M + N_{max}}{N_{max}}.
\]
The density matrix corresponding to this space has a block-matrix structure: the diagonal blocks correspond to mixtures with a fixed number of photons, while the off-diagonal blocks represent coherences between different photon number subspaces. Even though linear-optical elements cannot create superpositions of different photon numbers (off-diagonal blocks), we will consider them for the sake of generality, since such superpositions can be prepared before the interferometer and transmitted through it \cite{trivedi2020generation}. Equation (\ref{eq:rho_struct}) illustrates the block structure of the density matrix written in the basis of $\mathcal{H}_M$.
\begin{widetext}
\begin{equation}
    \rho_M = \begin{pmatrix}
        \boxed{\mathcal{H}^{(0)}_M\otimes \mathcal{H}^{(0)}_M} & \boxed{\mathcal{H}^{(0)}_M\otimes \mathcal{H}^{(1)}_M} & \dots & \boxed{\mathcal{H}^{(0)}_M\otimes \mathcal{H}^{(N_{max})}_M} \\
        \boxed{\mathcal{H}^{(1)}_M\otimes \mathcal{H}^{(0)}_M} &  \boxed{\mathcal{H}^{(1)}_M\otimes \mathcal{H}^{(1)}_M} &\dots&  \boxed{\mathcal{H}^{(1)}_M\otimes \mathcal{H}^{(N_{max})}_M}\\
        \vdots & \vdots & \ddots & \vdots\\
        \boxed{\mathcal{H}^{(N_{max})}_M\otimes \mathcal{H}^{(0)}_M} &  \boxed{\mathcal{H}^{(N_{max})}_M\otimes \mathcal{H}^{(1)}_M} & \dots &  \boxed{\mathcal{H}^{(N_{max})}_M\otimes \mathcal{H}^{(N_{max})}_M}
    \end{pmatrix}
\label{eq:rho_struct}
\end{equation}
\end{widetext}

\subsection{State Labeling and Tensor Product}

States in the basis are numbered in order of increasing photon number. For a fixed photon number $n$, states are ordered lexicographically. For example, for $M = 3$, $n = 2$, the states are ordered as:
\begin{align*}
&1)\ket{0,0,2},\quad 2)\ket{0,1,1},\quad 3)\ket{0,2,0}, \\& 4)\ket{1,0,1},\quad 5)\ket{1,1,0},\quad 6)\ket{2,0,0}.
\end{align*}
The tensor product operation in Fock space requires caution due to the non-fixed number of particles: when combining two systems with a variable number of photons, the total space also decomposes into a sum over possible total photon numbers. For instance, $\ket{1}$ and $\ket{0,1}$ are the first states in their respective bases (one photon in one mode and one photon in two modes), however, $\ket{1} \otimes \ket{0,1}$ has index 4 in the basis of 2 photons in 3 modes. Therefore, when taking the tensor product of states from two subsystems, the resulting matrix elements must be carefully renumbered.

\subsection{Evolution Operator}

The unitary evolution operator $\tilde{U}$, acting in Fock space, has a block-diagonal structure, where each block $\tilde{U}^{(n)}$ acts in the space with a fixed photon number $n$:
\begin{equation}
    \tilde{U} = \begin{pmatrix}
        1 & 0  & \dots & 0\\
        0 &  \boxed{\tilde{U}^{(1)}} &\dots &  0\\
        \vdots &  \vdots & \ddots & \vdots \\
        0& 0 & \cdots & \boxed{\tilde{U}^{(N_{max})}}
    \end{pmatrix}.
\end{equation}
The block-diagonal structure arises because the unitary matrix cannot induce transitions between states with different photon numbers.
Its elements are expressed through the permanents of submatrices of $U$:
\[
\tilde{U}^{(n)}_{\vec{i}, \vec{j}} = \frac{\mathrm{perm}(U_{\vec{i}}^{\vec{j}})}{\sqrt{\vec{i}!\vec{j}!}},
\]
where $\vec{i}$ and $\vec{j}$ are multi-indices describing the input and output configurations, $\vec{i}! = \prod_{k = 1}^M i_k!$, and $U_{\vec{i}}^{\vec{j}}$ is the submatrix of $U$ selected according to the standard rules for boson sampling \cite{scheel2004permanents}.
\section{Simulation Methods}

\subsection{Interferometer Unfolding}
One approach to modeling is the explicit unfolding of the circuit in time (see Fig. \ref{fig:mapping}) which was implemented in work~\cite{biriukov2025experimental}.
This leads to a boson sampling problem with $N \times T$ photons in $(M-L) \times T + L$ modes, but it does not account for the internal structure of the interferometer, such as the fact that photons from the $k$-th iteration cannot be detected in the previous $M\times(k-1)$ modes. This approach quickly becomes computationally inefficient but is easily implemented using standard libraries for linear optical quantum computing \cite{heurtel2022perceval, he2025deepquantum}.
\subsection{Partial Density Matrix Method}
A more efficient approach is based on computing the partial density matrix in the looped modes at each iteration.\\
At the end of each time iteration, photons are detected in the subsystem of non-looped modes $M_1$. If we assume the detection result is unknown, then to determine the state that propagates into the looped modes for the next time iteration, it is necessary to calculate the overall output state of the interferometer in all $M$ modes, and then take the partial trace over the subsystem of modes $M_1$ to obtain the partial density matrix in the looped channels $M_2$.\\
The Hilbert space $\mathcal{H}^{(N)}_M$ of $N$ photons in $M = M_1+M_2$ modes has a complex structure:
\begin{equation*}
    \mathcal{H}^{(N)}_M = \bigoplus_{n=0}^N \mathcal{H}^{(n)}_{M_1} \otimes \mathcal{H}^{(N-n)}_{M_2},
\end{equation*}
therefore, to find the density matrix $\rho_{M_2}$ from the full density matrix $\rho_{M_1M_2}$, one must take the partial trace over all possible subspaces $\mathcal{H}^{(n)}_{M_1}$:
\begin{align*}
    \label{eq:part_trace}
    &\rho_{M_2} = \operatorname{Tr}_{M_1}(\rho_{M_1M_2}) =\\
    &\sum_{n=0}^N \sum_{i_n^{M_1}} \left[ \bra{i_n^{M_1}} \otimes I^{(N-n)}_{M_2} \right] \rho_{M_1M_2} \left[ \ket{i_n^{M_1}} \otimes I^{(N-n)}_{M_2} \right],
\end{align*}
where $i_n^{M_1}$ are the basis vectors of the space $\mathcal{H}^{(n)}_{M_1}$, and $I^{(N-n)}_{M_2}$ is the identity operator in the space $\mathcal{H}^{(N-n)}_{M_2}$.\\
The algorithm for computing the distribution at the $k$-th time iteration is then as follows:
\begin{algorithm}[H]
\caption{Partial Density Matrix Evolution}
\begin{algorithmic}
\State $\rho^{\text{loop}}_0 = \ketbra{0}{0}$
\For{$i = 0$ to $k-1$}
    \State $\rho^{\text{in}}_i = \rho^{\text{ext}} \otimes \rho^{\text{loop}}_i$
    \State $\rho^{\text{out}}_i = \tilde{U} \rho^{\text{in}}_i \tilde{U}^\dagger$
    \State $\rho^{\text{loop}}_{i+1} = \operatorname{Tr}_{M_1} (\rho^{\text{out}}_i)$ \Comment{Note: Trace is over $M_1$, the non-looped/detected modes}
\EndFor
\State $\rho^{\text{det}}_k = \operatorname{Tr}_{M_2}(\rho^{\text{out}}_{k-1})$ \Comment{Trace over $M_2$, the looped modes, to get state for detection}
\end{algorithmic}
\end{algorithm}

\noindent
The diagonal of $\rho^{\text{det}}_k$ gives the desired photon distribution in the observable modes. The complexity of this method grows linearly with the number of iterations: it is necessary to compute all required evolution operators $\tilde{U}$ once, and then, regardless of the iteration number, all computations reduce to matrix multiplication.
\subsection{Kraus Operators in Linear Optics}

\subsubsection{Kraus Operators for Unitary Evolution and Partial Trace}

Consider a system consisting of two subsystems: an "external" part $M_1$ and a "loop" part $M_2$. The overall input state is given as a tensor product:
\[
\rho_{\text{in}} = \rho_{\text{ext}} \otimes \rho_{\text{loop}},
\]
where $\rho_{\text{ext}}$ is the density matrix in the external modes, and $\rho_{\text{loop}}$ is the density matrix in the loop.

The evolution of the entire system is governed by a unitary transformation:
\[
\rho_{\text{out}} = \tilde{U} (\rho_{\text{ext}} \otimes \rho_{\text{loop}}) \tilde{U}^\dagger.
\]

To obtain the updated state in the loop, the partial trace is taken over the external modes $M_1$:
\[
\rho_{\text{loop}}' = \operatorname{Tr}_{M_1}(\rho_{\text{out}}).
\]

Assume first that the input to the external modes is a pure Fock state $\ket{n}$. Then:
\[
\rho_{\text{ext}} = \ket{n}\bra{n}, \quad \rho_{\text{in}} = \ket{n}\bra{n} \otimes \rho_{\text{loop}}.
\]

After evolution and the partial trace:
\[
\rho_{\text{loop}}' = \operatorname{Tr}_{M_2}\left[ \tilde{U} (\ket{n}\bra{n} \otimes \rho_{\text{loop}}) \tilde{U}^\dagger \right].
\]

Introduce an orthonormal basis $\{\ket{m}\}$ in the space $\mathcal{H}_{M_1}$. The partial trace can then be rewritten as:
\[
\rho_{\text{loop}}' = \sum_{m \in \mathcal{H}_{M_1}} \bra{m} \tilde{U} (\ket{n}\bra{n} \otimes \rho_{\text{loop}}) \tilde{U}^\dagger \ket{m}.
\]

Rewrite this as a sum over Kraus operators:
\[
\rho_{\text{loop}}' = \sum_{m \in \mathcal{H}_{M_1}} K_m^{(n)} \rho_{\text{loop}} (K_m^{(n)})^\dagger,
\]
where
\[
K_m^{(n)} = \bra{m} \tilde{U} \ket{n} = \bra{m} \otimes \mathbb{I}_{M_2} \cdot \tilde{U} \cdot \ket{n} \otimes \mathbb{I}_{M_2}.
\]

Now consider an arbitrary density matrix $\rho_{\text{ext}}$ with spectral decomposition:
\[
\rho_{\text{ext}} = \sum_j \lambda_j \ket{\psi_j} \bra{\psi_j}, \quad \ket{\psi_j} = \sum_\alpha c_\alpha^{(j)} \ket{\alpha}.
\]

The result of the partial trace is then:
\[
\rho_{\text{loop}}' = \sum_{j} \sum_{m \in \mathcal{H}_{M_2}} K_m^{(\psi_j)} \rho_{\text{loop}} (K_m^{(\psi_j)})^\dagger,
\]
where
\[
K_m^{(\psi_j)} = \sqrt{\lambda_j}\bra{m} \tilde{U} \ket{\psi_j} = \sqrt{\lambda_j}\sum_{\alpha} c_\alpha^{(j)} \bra{m} \tilde{U} \ket{\alpha}.
\]

Thus, each input vector $\ket{\psi_j}$ generates a set of Kraus operators $K_m^{(\psi_j)}$ acting in the space of the looped modes $M_1$. This formalizes the evolution via a quantum channel:
\[
\mathcal{E}(\rho_{\text{loop}}) = \sum_i K_i \rho_{\text{loop}} K_i^\dagger,
\]
with a specific set $\{K_i\}$ constructed from the input state and the unitary operator $\tilde{U}$.\\
The method for calculating the distribution in the detectable modes then reduces to the following algorithm.

\begin{algorithm}[H]
\caption{Evolution Algorithm via Kraus Operators}
\begin{algorithmic}
\State $\rho^{\text{loop}}_0 \gets 0$
\For{$i = 0$ to $k-1$}
    \State $\rho^{\text{loop}}_{i+1} \gets \mathcal{E} (\rho^{\text{loop}}_i)$
\EndFor
\State $\rho^{\text{in}}_k \gets \rho^{\text{ext}} \otimes \rho^{\text{loop}}_k$
    \State $\rho^{\text{out}}_k \gets \tilde{U} \rho^{\text{in}}_k \tilde{U}^\dagger$
    \State $\rho^{\text{det}}_k \gets \operatorname{Tr}_{M_2}(\rho^{\text{out}}_{k})$
\end{algorithmic}
\end{algorithm}
If the Kraus operators for the joint unitary evolution and subsequent partial trace over a subset of modes act not in the infinite-dimensional Fock space but in a truncated one, their sum $\sum_nK_n^\dagger K_n$ will not equal the identity operator. This is because transitions from a state with $N_{\text{max}}$ photons to a state with, for example, $N_{\text{max}}+1$ photons will not be accounted for when the state space is truncated. However, in practice, this limitation can be circumvented by taking $N_{\text{max}} \geq N_{env}k$, where $N_{env}$ is the maximum number of photons in the input state $\rho_{ext}$ entering the non-looped modes, and $k$ is the number of considered time iterations. Then, for all states with a smaller number of photons, all possible transitions will be accounted for, and the Kraus operator formalism will work correctly.
\subsubsection{Photon Loss as a Quantum Channel}

Losses in photonic circuits are naturally modeled as interaction with an external environment (reservoir), to which the system is coupled via beam splitters. Let each mode experience losses with a transmission coefficient $T$ (and accordingly, reflection $1 - T$).

To model this situation, an auxiliary system (reservoir) is introduced, initially in the vacuum state $\ket{0}_r$. Let the state of the input mode be a Fock state $\ket{n}_s$. The combined initial state is:
\[
\ket{n}_s \otimes \ket{0}_r.
\]

Losses are modeled by the action of a unitary transformation $U_{BS}$, implemented by a beam splitter, which couples the input modes to the reservoir modes as follows:
\begin{align*}
&\begin{pmatrix}
    a'_s\\
    a'_r
\end{pmatrix} = U_{BS}\begin{pmatrix}
    a_s\\
    a_r
\end{pmatrix} \\
&U_{BS} =
\begin{pmatrix}
\sqrt{T} & \sqrt{1 - T} \\
-\sqrt{1 - T} & \sqrt{T}
\end{pmatrix},
\end{align*}
where $T$ is the power transmission coefficient of the input mode.
After the action of $U_{BS}$, the system transitions into an entangled state between the signal and the reservoir. If no detection is performed in the reservoir, the state in the signal modes is determined by taking the partial trace over the reservoir mode. The resulting mixed state in the signal mode is:
\[
\rho' = \operatorname{Tr}_r \left[\tilde{U}_{BS} \left(\ket{n}_s \bra{n} \otimes \ket{0}_r \bra{0}\right) \tilde{U}_{BS}^\dagger\right].
\]

This expression can be cast into the form of a quantum channel with Kraus operators. Let's find the explicit form of these operators.

It is known that
\[
\tilde{U}_{BS} \ket{n}_s \otimes \ket{0}_r = \sum_{k=0}^n
\sqrt{\binom{n}{k} T^{k} (1 - T)^{(n - k)}} \ket{k}_s \otimes \ket{n - k}_r.
\]

Taking the partial trace over the environment mode:
\[
\rho_s
' = \sum_{k=0}^n \binom{n}{k} T^k (1 - T)^{n - k} \ket{k}_s \bra{k}.
\]

This corresponds to the action of a quantum channel with Kraus operators:
\[
K_{k} = \sum_{n=0}^{\infty}\sqrt{\binom{n}{k} T^{k}(1 - T)^{(n-k)}} \ket{k}\bra{n}.
\]

The corresponding quantum channel in a space with a maximum photon number $N_{max}$ is written as:
\[
\mathcal{L}(\rho) = \sum_{k=0}^{N_{max}} K_{n} \rho K_{n}^\dagger.
\]
It is easy to verify that $\sum_kK_k^\dagger K_k = I$, so this is indeed a quantum channel describing photon losses in a given mode. \\
For a multimode circuit, the tensor product of single-mode loss operators is applied for each mode separately, taking into account the renumbering of states when taking the tensor product in the linear optical system. Thus, losses are represented as a quantum channel, and the overall quantum channel of unitary evolution with losses and partial trace is represented as the sequential action of channels, for example, in the case of losses only at the input to the looped modes: $\mathcal{E}\left(\mathcal{L}(\rho^{\text{loop}})\right) \equiv \mathcal{E'}(\rho^{\text{loop}})$.
\section{Stationary State of the System with Feedback}
\label{station_state_general}
In boson sampling systems with optical feedback, photons exiting from a subset of modes are fed back into the input of the circuit. If a fixed external state is also injected into the system at each time step, the system implements a discrete-time dynamics with iterative evolution of the state in the looped modes.

The state in the looped modes at the $i$-th iteration is expressed through its state at the previous iteration:
\[
\rho^{\text{loop}}_{i} = \mathcal{E} (\rho^{\text{loop}}_{i-1}).
\]
By iteratively applying the quantum channel, the state in the looped modes at any iteration can be expressed in terms of the initial vacuum state:
\begin{equation*}
    \rho^{\text{loop}}_{i} = \mathcal{E}^{(i)}(\ketbra{0}{0})
\end{equation*}
Looking at this equation one may suggest that over time a stationary state will be established in the looped modes:
\begin{equation*}
    \rho^{\text{loop}}_{stat} = \lim_{i\to\infty} \mathcal{E}^{(i)}(\ketbra{0}{0})
\end{equation*}

The stationary state of the system is defined as a state $\rho_{\text{loop}}^{\text{stat}}$ that remains unchanged under the action of the channel, i.e., it is its fixed point:
\[
\mathcal{E}(\rho_{\text{loop}}^{\text{stat}}) = \rho_{\text{loop}}^{\text{stat}}.
\]
According to Schauder's fixed point theorem \cite{schauder1930fixpunktsatz}, any continuous mapping of a convex compact set into itself in a Hilbert space has a fixed point. The quantum channel $\mathcal{E}$, acting on the set of density matrices, satisfies the conditions of the theorem. However, Schauder's theorem does not imply that the fixed point is unique. For the fixed point to be unique, the quantum channel must be strictly contracting throughout the entire Hilbert space $\mathcal{H}$, i.e.:
\begin{equation*}
    D\left(\mathcal{E}(\rho), \mathcal{E}(\sigma)\right) < D\left(\rho, \sigma\right), \quad \forall \rho,\sigma\in \mathcal{H}
\end{equation*}
where $D\left(\rho, \sigma\right)$ is the trace distance between the density matrices. \\
Since taking the partial trace is a strictly contracting mapping for the vast majority of photonic state pairs at the output of the linear optical interferometer (as they are highly entangled in terms of photon number across the detectable and looped mode subsets), some information about the state will inevitably be lost during the partial trace, which generally leads to a decrease in the trace distance, so the stationary state of this system will be unique.\\ 
The stationary state may not be unique if the interferometer matrix has a special structure, for example, if there is no coupling between the looped and external modes. In this case, any initial state in the looped modes will be stationary.\\
More specific requirements on the interferometer's transfer matrix will be formulated in subsequent sections. It turns out that for random interferometer transfer matrices the stationary state will be unique with probability 1.\\
The existence of a stationary state implies that photons will not accumulate indefinitely in the looped channels, meaning that for any number of considered iterations, there exists an saturated $N_{\text{max}}$ that will be sufficient for a complete description of the Hilbert space of the state in the looped modes. It will be shown later how to computationally obtain an estimate for $N_{\text{max}}$.

\subsection{Numerical Calculation of the Stationary State}

To numerically find $\rho_{\text{loop}}^{\text{stat}}$ in a finite-dimensional Fock subspace, it is necessary to construct the superoperator $\mathcal{E}$ in matrix form. For this, we vectorize the density matrix:
\[
\mathrm{vec}(\rho) \in \mathbb{C}^{d^2},
\]
where $d$ is the dimension of the Hilbert space of the looped modes.

The channel $\mathcal{E}$ acts linearly on the vectorized state:
\[
\mathrm{vec}(\rho_{i+1}) = G \cdot \mathrm{vec}(\rho_i),
\]
where $G$ is the superoperator matrix.

The stationary state corresponds to the eigenvector of $G$ with eigenvalue 1:
\[
G \cdot \mathrm{vec}(\rho^{\text{stat}}) = \mathrm{vec}(\rho^{\text{stat}}).
\]

Numerically, this leads to the problem of finding the principal eigenvector of a non-degenerate matrix. After finding $\mathrm{vec}(\rho)$, it must be reshaped back into a matrix
\[
\rho^{\text{stat}} =\mathrm{vec^{-1}}(\mathrm{vec}(\rho^{stat})).
\]

Thus, the task of computing the stationary state reduces to linear algebra — finding the eigenvector of the superoperator defining the quantum channel, with an eigenvalue equal to 1.

\section{Correlation Tensor Formalism and Density Matrix Reconstruction}

For analyzing quantum states, it is convenient to use correlation functions $C^{(k,l)}_{i_1 \dots i_k, j_1 \dots j_l} = \left\langle a_{i_1}^\dagger \dots a_{i_k}^\dagger a_{j_1} \dots a_{j_l} \right\rangle$, which define the statistical moments of the photon distribution. These tensors allow one to avoid storing the full density matrix in the Fock basis, which is especially important for large dimensions. Moreover, these tensors transform in a well-defined way under the unitary matrix of the interferometer and can be used to validate quantum calculations. Furthermore, they can be used to reconstruct the stationary state of the system and to formulate a theorem on the existence of a unique stationary state. Let us first consider a special case and then generalize it.

\subsection{First and Second Order Tensors}

Define the first and second order tensors:
\[
C^{(1)}_i = \langle a_i \rangle, \qquad C^{(1,1)}_{ij} = \langle a_i^\dagger a_j \rangle.
\]

Under the action of a unitary transformation $U$, the creation and annihilation operators transform as:
\[
a_i \rightarrow \sum_j U_{ij} a_j.
\]
Then the tensors transform as:
\begin{equation}
C'^{(1)} = U C^{(1)}, \qquad C'^{(1,1)} = V C^{(1,1)} V^\dagger,
\label{eq:cor_transform}
\end{equation}
where $V = U^*$.
In the case of feedback, the modes are split into external ($E$) and looped ($L$). Then $U$ has a block structure:
\[
U = \begin{pmatrix}
U_{EE} & U_{EL} \\
U_{LE} & U_{LL}
\end{pmatrix}.
\]
Similarly to how the matrix $U$ is split into blocks, the correlation tensors can also be split into blocks. We can assume that for tensors of all orders, the blocks containing only indices $E$ are known, e.g., $C^{(1)}_E, C^{(1,1)}_{EE}$, etc. To start, let's find the stationary value of the tensor $C^{(1)} = \left(C^{(1)}_E C^{(1)}_L\right)^T$. Assuming from the stationarity condition that $C'^{(1)}_L = C^{(1)}_L$, we solve the system of equations \ref{eq:cor_transform} for this block:
\begin{equation}
    C^{(1)}_L = (I-U_{LL})^{-1}U_{LE}C^{(1)}_E
\end{equation}
The tensor $C^{(1,1)}_{LL}$ of the looped modes in the stationary regime satisfies the equation:
\begin{align}
C^{(1,1)}_{LL} = V_{LE} C^{(1,1)}_{EE} V_{LE}^\dagger + V_{LL} C^{(1,1)}_{LL} V_{LL}^\dagger\\ + V_{LE} C^{(1,1)}_{EL} V_{LL}^\dagger + V_{LL} C^{(1,1)}_{LE} V_{LE}^\dagger.
\label{stat_cor}
\end{align}
In this expression, $C_{EE}$ is determined from the input state in the external modes, and the blocks $C^{(1,1)}_{LE}, C^{(1,1)}_{EL}$ are, firstly, related by the definition $C^{(1,1)}_{LE}={C^{(1,1)}_{EL}}^\dagger$, and secondly, can be determined from the first-order tensors. \\
Consider the general second-order tensor:
\[
C^{(1,1)}_{ij} = \langle a_i^\dagger a_j \rangle,
\]
where $i, j$ run over all modes — both external $E$ and looped $L$. Then the matrix $C^{(1,1)}$ is split into blocks:
\[
C^{(1,1)} =
\begin{pmatrix}
C_{EE} & C_{EL} \\
C_{LE} & C_{LL}
\end{pmatrix}.
\]
The initial state of the system is given as a tensor product:
\begin{equation}
\rho = \rho_E \otimes \rho_L,
\label{eq:input_rho}
\end{equation}
since before interaction through the interferometer, the subsystems of external and looped modes are not correlated. This implies that for any operators $A$ acting on $E$ and $B$ acting on $L$:
\begin{equation}
\langle A \otimes B \rangle = \langle A \rangle_{\rho_E} \cdot \langle B \rangle_{\rho_L}.
\label{eq:factorization}
\end{equation}
Apply this to the correlator \( \langle a_i^\dagger a_j \rangle \), where $i \in E$, $j \in L$:
\[
\langle a_i^\dagger a_j \rangle = \langle a_i^\dagger \rangle \cdot \langle a_j \rangle.
\]
Then
\[
(C^{(1,1)}_{EL})_{ij} = \langle a_i^\dagger \rangle \cdot \langle a_j \rangle = (C^{(1)}_E)^\dagger \otimes C^{(1)}_L,
\]
where \( C^{(1)}_E \) is the vector of averages over external modes, and \( C^{(1)}_L \) is over looped modes.
That is:
\[
C^{(1,1)}_{LE} = (C^{(1)}_E)^* \cdot (C^{(1)}_L)^T.
\]
Thus, we have defined all blocks of the correlation matrix except $C^{(1,1)}_{LL}$. We determine it by solving the system of equations (\ref{stat_cor}) using vectorization:
\begin{align}
&\mathrm{vec}(C^{(1,1)}_{LL}) = \left( I - V_{LL}^* \otimes V_{LL} \right)^{-1}\mathrm{vec}(S), \\
&S = V_{LE} C^{(1,1)}_{EE} V_{LE}^\dagger + V_{LE} C^{(1,1)}_{EL} V_{LL}^\dagger + V_{LL} C^{(1,1)}_{LE} V_{LE}^\dagger.
\label{eq:stat_cor}
\end{align}

\subsection{Tensors of Arbitrary Order}
In the previous section it was shown that one can use the correlation tensors of rank 1 to compute the stationary tensors of rank 2. Now we will generalize this procedure to tensors of order $(k,l)$:
\[
C^{(k,l)}_{i_1 \dots i_k, j_1 \dots j_l} = \left\langle a_{i_1}^\dagger \dots a_{i_k}^\dagger a_{j_1} \dots a_{j_l} \right\rangle.
\]

Under the action of $U$ this tensor is transformed in the following manner:
\[
C^{(k,l)'} = V^{\otimes k} C^{(k,l)} (V^\dagger)^{\otimes l}.
\]

The stationarity of the state in the looped modes is reflected in the system of equations:
\[
C_L^{(k,l)} = \big[V^{\otimes k} C^{(k,l)} (V^\dagger)^{\otimes l}\big]_L + S^{(k,l)},
\]
where the index $L$ denotes that we consider only those elements of the tensor whose indices all belong to the subset of looped modes, and $S^{(k,l)}$ is a tensor that includes transformed correlations between external modes $E$ (known from the input density matrix), as well as transformed correlations between external and looped modes, which factorize according to equation \ref{eq:factorization} into a product of tensors of lower orders, defined in previous iterations. Therefore, this term is considered known, and the remaining unknown terms can be found using vectorization and solving the system of equations:
\begin{equation}
    \mathrm{vec}[C_L^{(k,l)}] = \big[I_{L^{k+l}\times L^{k+l}} - \bigotimes_kV_{LL}^*\bigotimes_lV_{LL}\big]^{-1}\mathrm{vec}[S^{(k,l)}],
\end{equation}
which is a generalization of equation \ref{eq:stat_cor}. This system of equations will have a solution for all orders if the spectrum of the matrix $V_{LL}$ does not contain eigenvalues lying on the unit circle, regardless of the input state. From this, we can formulate a theorem on the existence of a stationary state in an interferometer with optical feedback:

\paragraph*{\textbf{Theorem on the Existence of a Stationary State.}} Consider an $M$-mode linear optical system with a unitary transfer matrix $U$. The last $L$ output modes of this system are looped back to the same last $L$ input modes. Let $U_{LL}$ be the block of the matrix $U$ characterizing the transition amplitudes from input looped modes to output looped modes, and let the same quantum state $\rho$ be periodically injected into the non-looped input modes of the system. If the spectral radius of $U_{LL}$ is strictly less than 1, then, regardless of the initial state, a stationary state determined solely by the transfer matrix $U$ and $\rho$ will be established in the looped modes over time. \\

This theorem gives the quantitative criterion on the interferometer's transfer matrix for the stationary state to exist. This completes the statement from Section \ref{station_state_general}.\\

Note that losses are incorporated into this formalism even more simply than into the Kraus operator formalism. Namely, it is necessary to replace the unitary matrix $U$ with the matrix $M = T_{out}UT_{in}$, where $T_{in}, T_{out}$ are diagonal matrices of amplitude transmission coefficients for the input and output modes, respectively. That is, the computational complexity of the method does not change when accounting for losses, unlike standard boson sampling calculation methods where losses are modeled by additional environment modes into which radiation is diverted.
\subsection{Recursive Algorithm for Tensor Computation}

The construction of correlation tensors in the looped modes is performed iteratively. For a tensor of order $(k, l)$, we use already calculated tensors of lower order. Thus, knowing the input density matrix $\rho_E$ and the interferometer's transfer matrix $U$, one can calculate the average values of any correlation tensors in the stationary state. The calculation algorithm is summarized in Algorithm \ref{alg:cor_recurs}.
\begin{algorithm}[H]
\caption{Recursive Computation of Correlation Tensors}
\begin{algorithmic}[1]
\For{$n = 1$ to $N_{\text{max}}$}
\For{$m = 0$ to $n$}
    \ForAll{multi-indices $\vec{i}, \vec{j}$ on the looped modes $L$}
        \State Compute the right-hand side $S^{(n,m)}_{\vec{i}, \vec{j}}$
    \EndFor
    \State Solve:
        \State $C_L^{(n,m)} = \big[V^{\otimes k} C^{(n,m)} (V^\dagger)^{\otimes l}\big]_L + S^{(n,m)},$
\EndFor
\EndFor
\end{algorithmic}
\label{alg:cor_recurs}
\end{algorithm}

\subsection{Reconstructing the Density Matrix from Correlation Tensors}
Assume that correlation tensors up to order $n \leq 2N_{\text{max}}$ are known. If we assume that the final state of the system can be accurately approximated by a state in the basis from 0 to $N_{\text{max}}$ photons in L modes, then these correlation tensors can be used to reconstruct the density matrix, since the action of more than $N_{max}$ annihilation operators on any state in this basis will yield 0. $N_{\text{max}}$ can be determined from second-order tensors $\braket{a^\dagger_ia_i}$, which give the average photon numbers $\braket{n_i}$ in the looped modes, and fourth-order tensors $\braket{a^\dagger_ia^\dagger_ia_ia_i}$, from which the variance of the photon number $\braket{(\Delta n_i)^2}$ in the looped modes can be determined. Then one can set, for example, $N_{max} = \sum_i (n_i + 3\sqrt{\braket{(\Delta n_i)^2}})$ (the three-sigma rule). All higher-order tensors will have small values. If it's not the case, like for thermal state which has mean value of correlators equal $n!n_{mean}^n$, than one should proceed calculations until the values are negligible. Now let's consider possible reconstruction strategies.\\
The stationary density matrix is written in the basis of the space $\mathcal{H}_L = \bigoplus_{n=0}^{N_{\text{max}}}\mathcal{H}^{(n)}_L$, $\dim \mathcal{H}_L = \begin{pmatrix}
    L+N_{max}\\
    N_{max}
\end{pmatrix}$:
\begin{equation*}
    \rho_{stat} = \sum_{n,m = 1}^{\dim \mathcal{H}_L} \rho_{nm}\ketbra{n}{m} = \sum_{n,m = 1}^{\dim \mathcal{H}_L} \frac{\rho_{nm}}{\vec{n}!\vec{m}!}A_n^\dagger\ketbra{0}{0}A_m,
\end{equation*}
where $A_n = \bigotimes_i a_i^{n_i}$ is the annihilation operator for the state numbered $n$ in the basis of $\mathcal{H}_L$. Consider the average value of the string $A^\dagger_k A_l$:
\begin{align*}
    &C_{k,l} = \braket{A^\dagger_k A_l} = \sum_{n,m = 1}^{\dim \mathcal{H}_L} \frac{\bra{0}A_m A_k^\dagger A_l A_n^\dagger\ket{0}}{\vec{n}!\vec{m}!}\rho_{nm} = \\&\sum_{n,m = 1}^{\dim \mathcal{H}_L} B^{nm}_{kl}\rho_{nm}
\end{align*}
We obtain a system of linear equations connecting the correlation tensors and the elements of the density matrix via the tensor $B$, whose non-zero elements can be calculated in advance, knowing the order of the tensors in $\mathrm{vec}[C]$ and the basis:
\begin{equation}
    B\mathrm{vec}[\rho] = \mathrm{vec}[C].
    \label{eq:conv_optim}
\end{equation}
The matrix $B$ can be viewed as a generalization to linear optical systems of the protocol matrix from the field of quantum state tomography, and $C$ as a generalization of the probabilities of detecting a particular outcome in a certain basis.
This system of equations can be solved using convex optimization. However, this quickly becomes impractical as the dimension grows quickly. \\
An alternative way is to calculate the density matrix elements analytically. The system of equations \ref{eq:conv_optim} is upper triangular, since the contribution of $\rho_{nm}$ to the moment $C_{kl}$ is only possible if:
\begin{equation*}
    n_i \geq k_i, m_i \geq l_i, \quad \forall i\in 1,\dots,M
\end{equation*}
Consequently, if all moments up to a total rank $K\ge \sum_i n_i+\sum_i m_i$ are known, one can recursively express $\rho_{nm}$ in terms of other moments and already reconstructed elements with higher total photon numbers. A more detailed derivation is discussed in Appendix \ref{sec:reconstruction}. This method requires significantly fewer computational resources but is more sensitive to inaccuracies in determining $C_{kl}$. Furthermore, if not all moments with significant magnitude are known, this algorithm can lead to negative probabilities, necessitating projection onto the nearest density matrix. This cannot be achieved with convex optimization, as in that method the search is confined a priori to the space of density matrices.
\section{Numerical Modeling}
\subsection{Time to Reach Stationary State}
\begin{figure}
    \centering
    \includegraphics[width=1\linewidth]{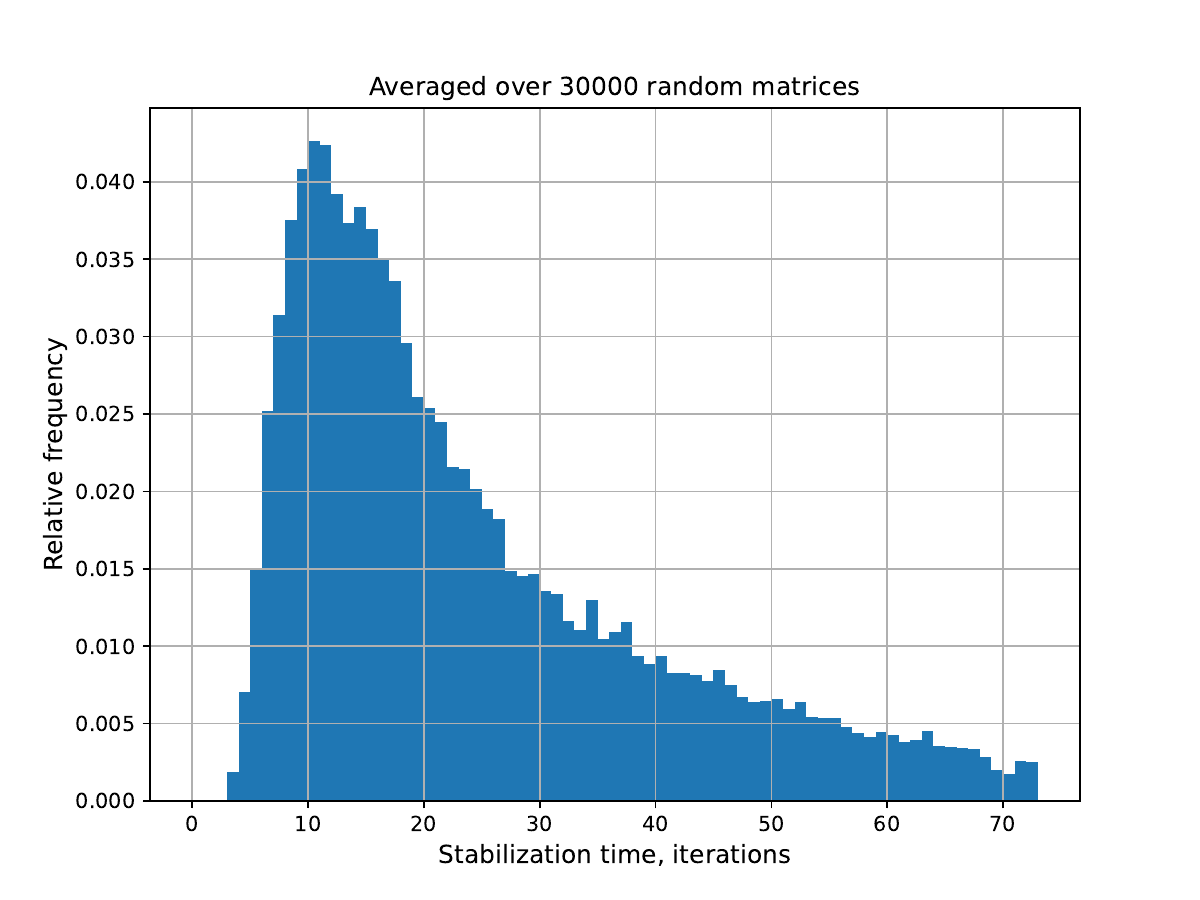}
    \caption{Distribution of system stabilization times for a two-mode system with one feedback channel in the absence of losses. The distribution is obtained from a sample of 30,000 random matrices.}
    \label{fig:stab_time}
\end{figure}

One of the quantities characterizing a boson sampler with optical feedback channels is the number of iterations required for a stationary distribution to be established in the looped modes. We will call this quantity the \textit{stabilization time} $\tau_{st}$. \\
$\tau_{st}$ depends on the interferometer matrix and the input state. Let us fix the input state as $\ket{1\dots1}$, i.e., a single photon in each non-looped mode. Figure \ref{fig:stab_time} shows the distribution of $\tau_{st}$ for 30,000 random unitary $2\times2$ matrices with one feedback channel in the absence of losses.
The graph shows that the distribution has a long right tail, which is mainly determined by matrices with small transition amplitudes between looped and non-looped modes, but overall, the values are clustered around 15 iterations. The stop condition was chosen as infidelity between true stationary state that was calculated using superoperator and the state obtained after subsequent application of Kraus operators is less than $10^{-6}$. 

\subsection{Photon Distribution in the Looped Modes}
\begin{figure*}[ht]
    \centering
    \includegraphics[width=1\linewidth]{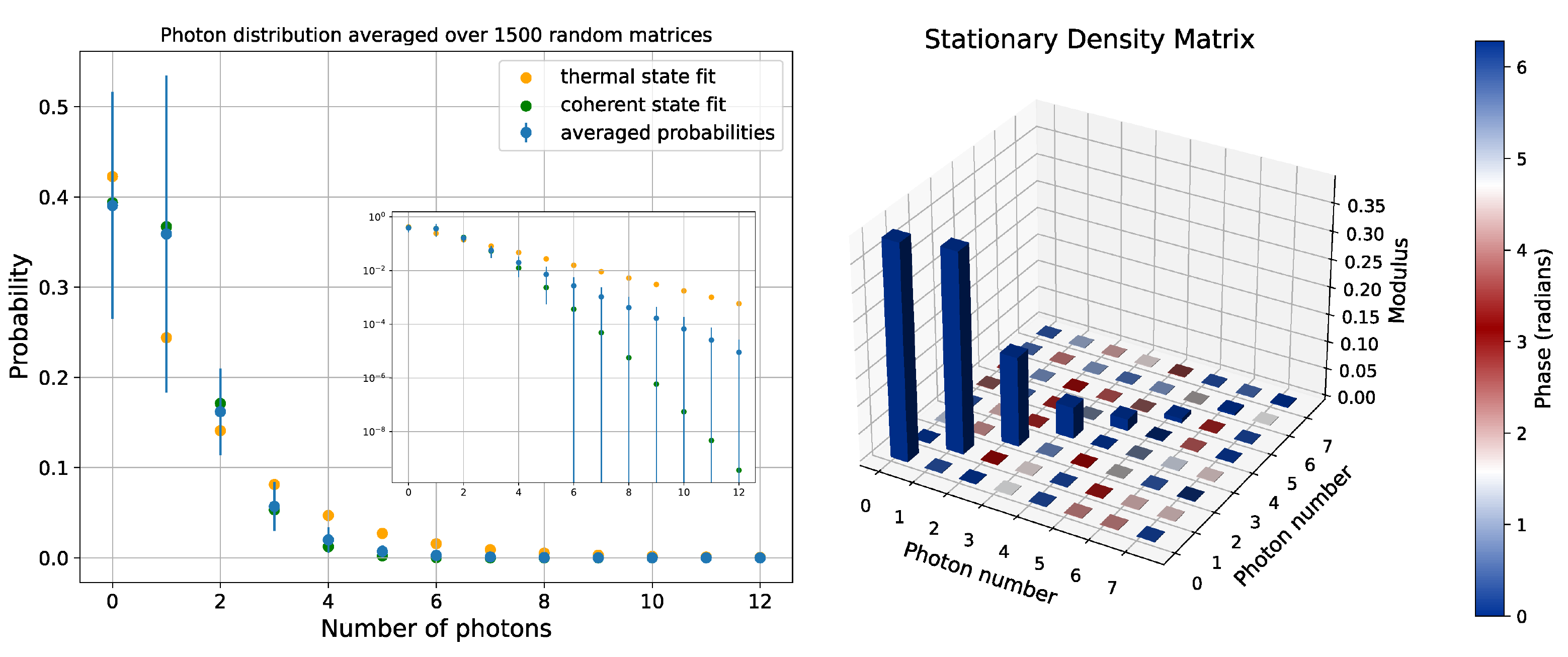}
    \caption{Characteristics of the stationary density matrix in the looped modes for a system with 2 modes, one of which is looped. (Left) Photon number distribution with the closest approximation by a thermal and a coherent state. (Right) View of the average stationary density matrix for this system. Bar height represents the modulus of matrix elements, color represents the phase. Since the stationary state is established in one mode, each cell corresponds to a specific photon number.}
    \label{fig:photon_distrib}
\end{figure*}

\begin{figure*}[ht]
    \centering
    \includegraphics[width=1\linewidth]{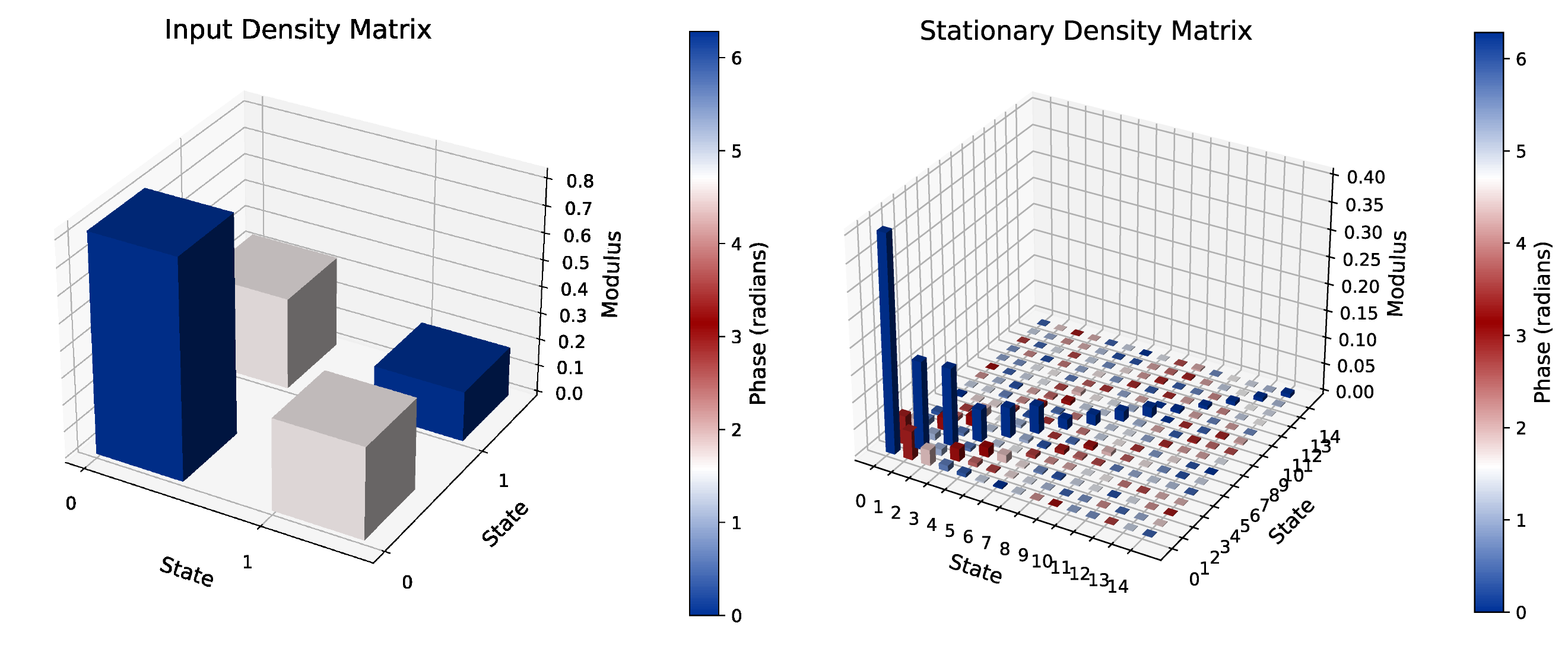}
    \caption{Three-mode system with two feedback channels and a random input state. (Left) Random input state in the basis from 0 to 1 photon in 1 mode. (Right) Stationary state, which no longer has a block-diagonal structure due to the presence of off-diagonal elements in the input density matrix. Unlike Fig. \ref{fig:photon_distrib}, the axes here represent not the photon number but the state indices in the basis $\mathcal{H}_L$.}
    \label{fig:random_input}
\end{figure*}

\begin{figure*}[ht]
    \centering
    \includegraphics[width=1\linewidth]{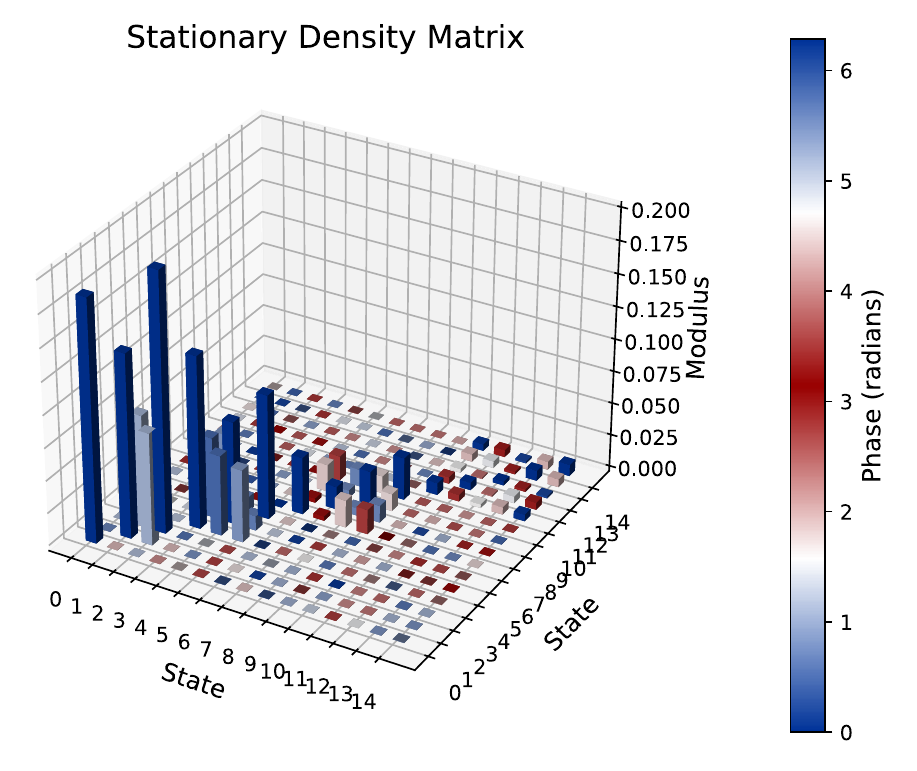}
    \caption{Stationary density matrix in a 3-mode system with 2 feedback channels. Unlike the case with one loop, this density matrix has off-diagonal elements corresponding to superpositions of states with fixed photon numbers. Unlike Fig. \ref{fig:photon_distrib}, the axes here represent not the photon number but the state indices in the basis $\mathcal{H}_L$.}
    \label{fig:two_loops}
\end{figure*}

The existence of a unique stationary state resembles a thermalization process, which is the stationary bosonic states will be occupied according to thermal state distribution with some effective temperature.  In all subsequent simulations, the Fock state $\ket{1\dots1}$ was injected into all non-looped modes at each iteration.\\
Figure \ref{fig:photon_distrib} shows the photon distribution in the stationary state for a two-mode system with one feedback channel, averaged over 1500 unitary matrices, as well as the average stationary density matrix computed using the formula $\rho^{(stat)}_{av} = \frac{1}{samples}\sum_{i=1}^{samples}\rho^{(stat)}_i$. This averaging is possible, since there are no off-diagonal elements in the density matrix of photons in one looped mode.  
This distribution was approximated both by a thermal distribution and by the photon distribution in a coherent state. The inset of Fig. \ref{fig:photon_distrib}, showing the same graph on a logarithmic scale, reveals that within the error margin, the photon statistics can be described using a coherent state; however, the discrepancy increases with the photon number. \\
Thus, the suggestion of system thermalization is not valid even in the simplest case. Increasing the number of non-looped modes does not lead to significant changes in photon statistics. The diagram showing the density matrix reveals a block-diagonal structure, where the blocks represent states with a fixed number of photons. This is because the initial state contained no superpositions of different photon numbers, and thus, using Kraus operators, transitions between states with different photon numbers cannot be generated. If a random Fock state is injected into the non-looped modes (see Fig. \ref{fig:random_input}), then superpositions of different photon numbers will also be observed in the stationary state.
If a second looped mode is added (see Fig. \ref{fig:two_loops}), off-diagonal elements appear in the stationary density matrices, indicating the presence of superpositions of photon numbers in the stationary states. \\
\subsection{Quality of Density Matrix Reconstruction from Stationary Tensors}
\begin{figure*}[ht]
    \centering
    \includegraphics[width=1\linewidth]{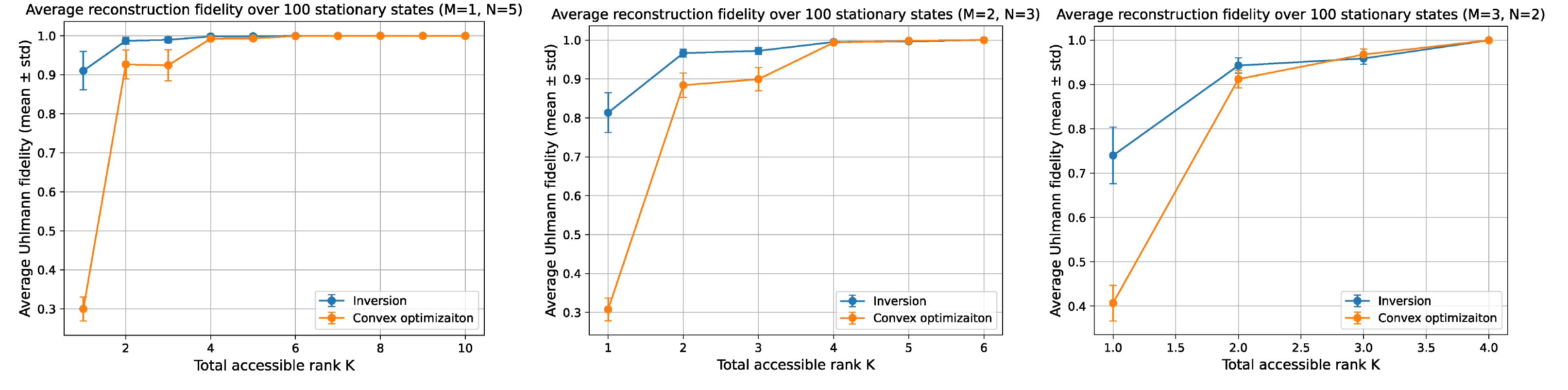}
    \caption{Accuracy of density matrix reconstruction from correlation tensors using analytical inversion (inversion) and convex optimization (convex optimization) for different numbers of looped modes $M$ and maximum photon numbers in the state $N$. It can be seen that in most cases, tensor inversion yields better results when all tensors are only partially determined.}
    \label{fig:rec_qual}
\end{figure*}

Two methods for reconstructing the stationary density matrix from correlation tensors were previously proposed: analytical inversion with projection onto the nearest density matrix, and convex optimization over the space of density matrices. It is interesting to see how the reconstruction quality depends on the maximum rank of the known tensors, since the number of components of the stationary tensors that need to be determined grows exponentially with rank.
 The graphs in Fig. \ref{fig:rec_qual} show the accuracy of state reconstruction, expressed as the Uhlmann fidelity between the exact stationary state, which was calculated using superoperator formalism and the state reconstructed using the two different methods. For this task, a boson sampler with high losses of 85\% was specifically chosen to artificially limit the maximum number of photons in the stationary state due to computational resource constraints. Surprisingly, the inversion method followed by projection onto the nearest density matrix almost always outperforms convex optimization, especially when only low-order tensors are known. The components of higher orders are small due to losses, so both methods quickly approach a fidelity close to 1.\\

\begin{figure*}
    \centering
    \includegraphics[width=1\linewidth]{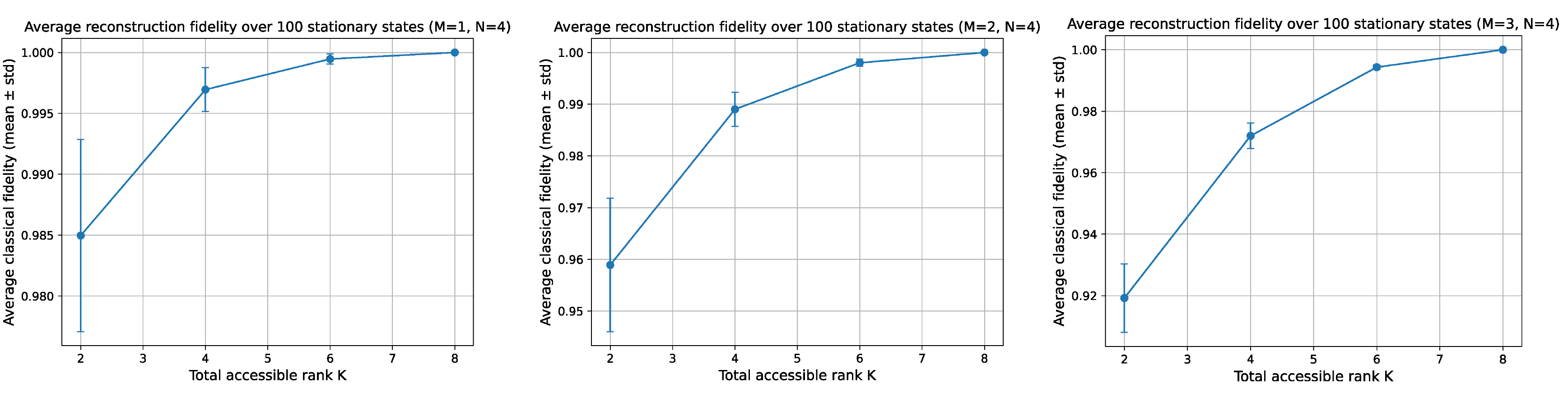}
    \caption{Results of reconstructing the probability distribution of the stationary state in the detectable modes for 100 random transfer matrices with different numbers of detectable modes $M$ and a fixed maximum number of photons in these modes $N$. The graphs show the dependence of the classical fidelity between the true probability distribution and the reconstructed one on the number of available symmetric tensors.}
    \label{fig:prob_rec}
\end{figure*}

\begin{figure}[ht]
    \centering
    \includegraphics[width=1\linewidth]{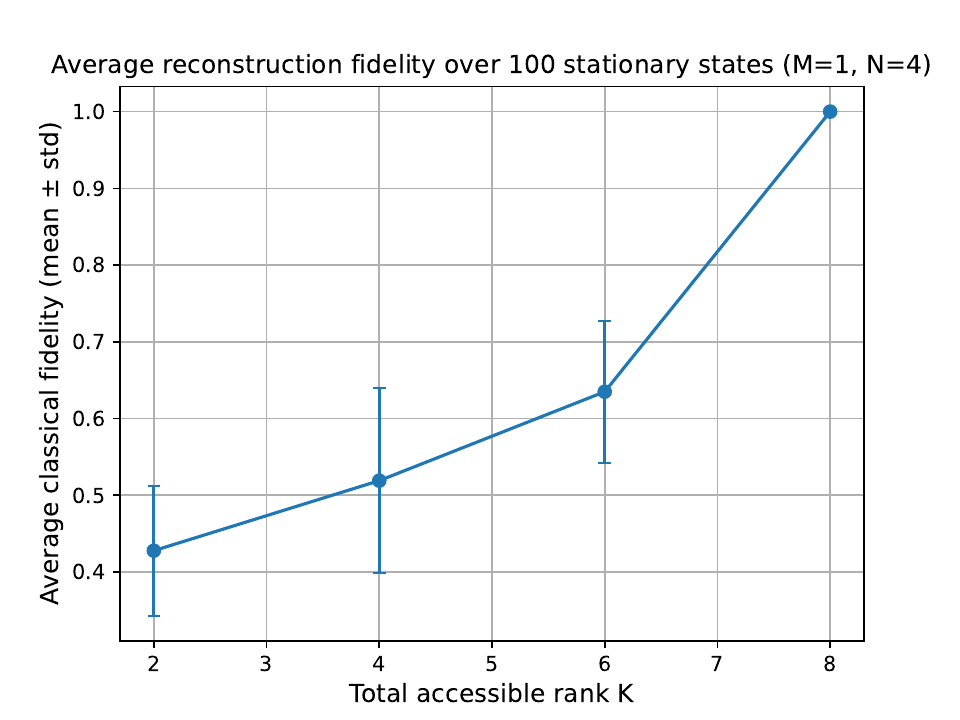}
    \caption{Results of reconstructing the probability distribution for 100 random density matrices written in the basis from 0 to 4 photons in 1 mode. The graphs show the dependence of the classical fidelity between the true probability distribution and the reconstructed one on the number of available symmetric tensors. Compared to Fig. \ref{fig:prob_rec}, the reconstruction quality has deteriorated because the contribution of multi-photon components to the moment values does not decrease with increasing photon number.}
    \label{fig:prob_rec_rand}
\end{figure}

To generate samples from the stationary distribution, it is not necessary to know the entire density matrix. It is sufficient to know its diagonal elements, which define the probability distribution of obtaining a particular outcome. This, in turn, can be calculated from tensors of rank type (s,s). If the maximum number of photons in the stationary state is $N_{max}$ and all tensors up to rank $(N_{max}, N_{max})$ are known, the distribution will be reconstructed exactly. Otherwise, when not all tensors are known, we obtain an approximate distribution. \\
The graphs in Fig. \ref{fig:prob_rec} show the dependence of the classical fidelity between two probability distributions $F(p,q) = \sum_i\sqrt{p_i q_i}$ on the maximum known tensor rank.
This was obtained as follows. First, the stationary state in the looped modes was found using an iterative tensor algorithm. Then, all stationary tensors of symmetric rank in the detectable modes were found by explicitly convolving all known tensor components at the input with the interferometer's transfer matrix. The diagonal of the density matrix was computed using the analytical method discussed in Appendix \ref{sec:reconstruction}. If the probability distribution obtained by this method was negative, which is quite likely if components with a large number of photons have significant probability, the distribution was projected onto the standard simplex. As in the previous case, the simulation was performed under conditions of high losses to limit the maximum number of photons in the output states.\\
The graph shows that as the number of modes increases, the quality of density matrix reconstruction with partially known tensors noticeably decreases. However, in all three cases, tensors of rank (3,3) are sufficient to achieve a fidelity greater than 99\%. With decreasing losses, the proportion of multi-photon components should increase, which could potentially worsen the reconstruction quality. To model this situation, one can consider not stationary density matrices, but random density matrices written in the same basis (see Fig. \ref{fig:prob_rec_rand}). In this case, even for one detectable mode, the fidelity remains extremely low until all necessary tensors become known.
\section{Discussion}
In the presented work, we developed a theoretical model of boson sampling with optical feedback, where a subset of modes input and output modes is looped, enabling more complex bosonic dynamics in spatio-temporal domain. The proposed calculation methods, ranging from simple interferometer unfolding to the formalism of Kraus operators and correlation tensors, conducted several numerical experiments at small scale to get the system's main features. In particular, the use of Kraus operators allows for a natural incorporation of losses and describes the evolution as a quantum channel leading to a unique stationary state for random unitary matrices. The correlation tensor formalism, in turn, simplifies the reconstruction of the density matrix and the analysis of correlations, which is especially useful for systems with a variable number of photons.
One of the key results is the theorem on the existence of a unique stationary state, based on the contracting properties of the partial trace and unitary evolution. This confirms that the system does not accumulate an infinite number of photons in the loop but reaches equilibrium, which simplifies numerical modeling: it is sufficient to choose a finite $N_{\max}$ based on an estimate of the average photon number.\\
The proposed architecture extends the capabilities of boson sampling through temporal dynamics, making it promising for quantum simulation. As discussed in the introduction, such a system can simulate molecular vibrational spectra \cite{huh2015boson,sparrow2018simulating}, non-adiabatic chemical dynamics \cite{fingerhuth2024quantum}, quantum walks in disordered systems \cite{caruso2016fast}, and bosonic thermodynamics \cite{arrazola2023quantum}. In these applications, the looped modes $L$ act as quantum memory storing the evolving state, while injection into the $M$ modes simulates external influences or discrete time steps. This circumvents the limitations of standard boson sampling, where the state is reinitialized each time, and allows for larger-scale simulations with fewer resources.\\
In terms of computational complexity, the simulation remains \#P-hard, but the proposed algorithms reduce the exponential dependence on the number of iterations by focusing on the evolution of the partial density matrix. Nevertheless, for large $M$ and $L$ (on the order of tens of modes), classical simulation becomes inefficient, highlighting the potential for demonstrating quantum supremacy. Including losses makes the model more realistic for photonic experiments but also introduces decoherence, which can weaken quantum effects; future work could investigate strategies for compensating losses, e.g., through post-selection \cite{aaronson2016bosonsamplingloss} or (heralded) signal amplification \cite{ralph2009nla,xiang2010heraldednla}.
Limitations of the approach include the approximation of a finite $N_{\max}$, the neglect of higher-order correlations in the tensor formalism, and the assumption of perfect unitarity. \\
Of particular interest is a new computational task naturally suggested by the proposed architecture: \emph{Stationary distribution boson sampling}. Concretely, one first considers the stationary state established in the looped modes $L$, $\rho_{\text{loop}}^{\text{stat}}$, and then defines the output distribution in the detectable modes obtained by interfering $\rho_{\text{loop}}^{\text{stat}}$ with the injected external state and tracing out the looped subsystem.

Determining the precise complexity class of this \emph{stationary-distribution sampling} problem is a subject of further research: unlike standard single-pass boson sampling, the distribution is defined implicitly via a fixed point of a quantum channel (equivalently, via the principal eigenvector of the corresponding superoperator). At the same time, it is clear that this task strictly generalizes the usual boson sampling problem: in the absence of feedback loops ($L=0$), it reduces to ordinary boson sampling, while for $L>0$ it combines the hardness of linear-optical interference (through permanents) with the additional difficulty of finding (or characterizing) the stationary state.
\section{Conclusion}
In conclusion, this work proposes a model of boson sampling with optical feedback and develops methods for computing output distributions and the stationary state using Kraus operators and correlation tensors. We prove that, in most cases, the system admits a unique stationary state, which is fundamental for understanding the resulting dynamics.\\
The obtained results pave the way for compact optical schemes that may demonstrate quantum computational advantage with fewer resources and enable the study of open quantum systems within a boson-sampling-based architecture.

\section{Acknowledgement}
The work was supported by Russian Science Foundation grant 22-12-00353-$\Pi$ (https://rscf.ru/en/project/22-12-00353/).The work was supported by Rosatom in the framework of the Roadmap for Quantum computing (Contract № 868/1759-D dated 3 October 2025 and Contract №11-2025/1 dated 14 November 2025). Yu.A. B is grateful to the Russian Foundation for the Advancement of Theoretical Physics and Mathematics (BASIS) (Projects №24-2-10-57-1).

The authors thank S.A. Fldzhyan for insightful discussions and thorough paper review.  
\appendix

\begin{widetext}
\section{Analytical Computation of the Density Matrix from Correlation Tensors}
\label{sec:reconstruction}
\subsection{Single-Mode Case}
In the single-mode case, the standard formula for the action of ladder operators on Fock states
\begin{equation}
a^\dagger{}^s a^r \ket{n} =
\begin{cases}
\sqrt{\dfrac{n!}{(n-s)!}}\,\sqrt{\dfrac{(n-s+r)!}{(n-s)!}}\;\ket{n-s+r}, & n\ge s,\\[6pt]
0,& n<s
\end{cases}
\end{equation}
yields the matrix elements
\begin{equation}
\bra{n}a^\dagger{}^s a^r\ket{m}
= \delta_{n-s,\, m-r}\,\sqrt{\frac{n!}{(n-s)!}}\,\sqrt{\frac{m!}{(m-r)!}}.
\end{equation}
Consequently, the moments
$C^{(s,r)}=\Tr(\rho\,a^\dagger{}^r a^s)$ are
\begin{equation}\label{eq:1mode_moment}
C^{(s,r)}
=\sum_{n,m}\rho_{n,m}\,\bra{m}a^\dagger{}^s a^r\ket{n}
=\sum_{n\ge s}\rho_{n,n-s+r}\,\sqrt{\frac{n!}{(n-s)!}}\,\sqrt{\frac{(n-s+r)!}{(n-s)!}}.
\end{equation}

\subsection{Multimode Case}
In the multimode case, the operators factorize over modes, hence
\begin{align}
\bra{\bs n}\Big(\prod_{i=1}^m a_i^{\dagger\, s_i} a_i^{\, r_i}\Big)\ket{\bs m}
&=\prod_{i=1}^m \bra{n_i} a_i^{\dagger\, s_i} a_i^{\, r_i} \ket{m_i}\\
&=\prod_{i=1}^m \delta_{n_i-s_i,\, m_i-r_i}\,
\sqrt{\frac{n_i!}{(n_i-s_i)!}}\,\sqrt{\frac{m_i!}{(m_i-r_i)!}}.
\end{align}
From this, the general moment is
\begin{equation}\label{eq:mu_rs_general}
C^{(\bs s,\bs r)}
=\sum_{\bs n,\bs m} \rho_{\bs n,\bs m}
\prod_{i=1}^m \delta_{n_i-s_i,\, m_i-r_i}\,
\sqrt{\frac{n_i!}{(n_i-s_i)!}}\,\sqrt{\frac{m_i!}{(m_i-r_i)!}}.
\end{equation}
The delta symbols enforce \emph{local matching} over modes:
$\bs m=\bs n-\bs s+\bs r$, and the sum runs over those $\bs n$ for which $\bs n\ge \bs s$ and $\bs m\ge \bs r$ component-wise.

\paragraph{System of Linear Equations.}
Denote the coefficient
\begin{equation}\label{eq:coeff_def}
B(\bs n,\bs m;\bs r,\bs s)\;=\;
\prod_{i=1}^m
\sqrt{\frac{n_i!}{(n_i-s_i)!}}\,
\sqrt{\frac{m_i!}{(m_i-r_i)!}}\;
\cdot\;
\prod_{i=1}^m \delta_{n_i-s_i,\, m_i-r_i}.
\end{equation}
Then \eqref{eq:mu_rs_general} is concisely written as
\begin{equation}\label{eq:lin_system}
C^{(\bs s,\bs r)}=\sum_{\bs n,\bs m} B(\bs n,\bs m;\bs r,\bs s)\,\rho_{\bs n,\bs m},
\qquad
\sum_i r_i=R,\;\sum_i s_i=S.
\end{equation}
This is an \emph{upper-triangular} system with respect to the partial order by total photons:
the contribution of $\rho_{\bs n,\bs m}$ to the moment $C_{\bs r,\bs s}$ is only possible if
\begin{equation}
\bs n \geq \bs r, \bs m \geq \bs s
\end{equation}
Consequently, if all moments up to a total rank $K\ge \sum_i n_i+\sum_i m_i$ are known, one can recursively express $\rho_{\bs n,\bs m}$ in terms of moments and already reconstructed elements with \lq\lq higher\rq\rq\ total photon sums.

% \section{Probabilities $p_{\bs n}$ (diagonal of $\rho$) from moments}

The probabilities $p_{\bs n}=\rho_{\bs n,\bs n}$ are extracted from \eqref{eq:lin_system} by taking $\bs r=\bs s=\bs n$:
\begin{equation}\label{eq:diag_basic}
C^{(\bs n,\bs n)}
=\sum_{\bs k,\bs \ell} B(\bs k,\bs \ell;\bs n,\bs n)\,\rho_{\bs k,\bs \ell}.
\end{equation}
Due to the matching conditions $k_i-n_i=\ell_i-n_i$, i.e., $\bs k=\bs n+\bs q$ and $\bs \ell=\bs n+\bs q$ with $\bs q\ge \bs 0$. Then
\begin{equation}\label{eq:diag_expand}
C^{(\bs n,\bs n)}
=\sum_{\bs q\ge \bs 0}
\Bigg[\prod_{i=1}^m \frac{(n_i+q_i)!}{n_i!}\Bigg]\rho_{\bs n+\bs q,\;\bs n+\bs q}.
\end{equation}
This leads to a top-down recursion over the total photon number $|\bs n|=\sum_i n_i$:
\begin{equation}\label{eq:prob_recur}
\rho_{\bs n,\bs n}
=C^{(\bs n,\bs n)}
-\sum_{\substack{\bs q\ge\bs 0\\ \bs q\neq \bs 0}}
\Bigg[\prod_{i=1}^m \frac{(n_i+q_i)!}{n_i!}\Bigg]\rho_{\bs n+\bs q,\;\bs n+\bs q},
\qquad |\bs n|\le N_{\max}.
\end{equation}
The recursion starts at the maximum photon number: for $|\bs n|=N_{\max}$ the second sum is empty, and
$\rho_{\bs n,\bs n}=C_{\bs n,\bs n}$. Then, decreasing $|\bs n|$, all probabilities are computed, including the vacuum $p_{\bs 0}$.

% \subsection{Reconstruction of All Matrix Elements $\rho_{\bs n,\bs m}$}

Now consider the general case $\bs n$ and $\bs m$ (possibly $\bs n\neq \bs m$).
From \eqref{eq:lin_system}, choose the moment with ranks $\bs r=\bs n$ and $\bs s=\bs m$.
Then the matching conditions \eqref{eq:coeff_def} require
$\bs k-\bs n=\bs \ell-\bs m=\bs q\ge \bs 0$,
and we obtain the \emph{triangular} expanded form
\begin{equation}\label{eq:offdiag_expand}
C^{(\bs n,\bs m)}
=\sum_{\bs q\ge \bs 0}
\Bigg[\prod_{i=1}^m \sqrt{\frac{(n_i+q_i)!}{n_i!}}\,
\sqrt{\frac{(m_i+q_i)!}{m_i!}}\Bigg]\;
\rho_{\bs n+\bs q,\;\bs m+\bs q}.
\end{equation}
From \eqref{eq:offdiag_expand}, a \emph{top-down recursion} immediately follows over the quantity
\begin{equation}
L(\bs n,\bs m):=\min\!\big(N_{\max}-|\bs n|,\; N_{\max}-|\bs m|\big),
\end{equation}
similar to \eqref{eq:prob_recur}:

\begin{equation}\label{eq:offdiag_final}
\rho_{\bs n,\bs m}
=C^{(\bs n,\bs m)}
-\sum_{\substack{\bs q\ge \bs 0\\ \bs q\neq \bs 0}}
\Bigg[\prod_{i=1}^m \sqrt{\frac{(n_i+q_i)!}{n_i!}}\,
\sqrt{\frac{(m_i+q_i)!}{m_i!}}\Bigg]\rho_{\bs n+\bs q,\;\bs m+\bs q}.
\end{equation}
Procedure: order the pairs $(\bs n,\bs m)$ by non-increasing $L(\bs n,\bs m)$, for each pair subtract the contribution of already found elements with larger $L$, and obtain $\rho_{\bs n,\bs m}$. The fact that $\rho_{\bs m,\bs n}=\overline{\rho_{\bs n,\bs m}}$ allows computing only the upper triangle (in a fixed lexicographical order of the basis).
\end{widetext}
\bibliography{bibliography}
\end{document}